\documentclass[thmsa,11pt]{article}
\usepackage{amssymb,amsmath}

\begin{document}

\title{\hfill {\small CECS-PHY-05/07\bigskip }\\
\textbf{Dynamics and BPS states of AdS}$_5$ \textbf{supergravity with a
Gauss-Bonnet term}}
\author{Olivera Mi\v{s}kovi\'{c}\,$^{*,\dagger}$, Ricardo Troncoso\,$^\ddagger$ and
Jorge Zanelli\,$^\ddagger\medskip \medskip $ \\
$^*${\small \emph{Departamento de F\'{i}sica, P. Universidad
Cat\'{o}lica de Chile, Casilla 306, Santiago 22, Chile}.}\\
$^\dagger${\small \emph{Instituto de F\'{i}sica, P. Universidad
Cat\'{o}lica de Valpara\'{\i}so, Casilla 4059, Valpara\'{\i}so, Chile}.}\\
$^\ddagger${\small \emph{Centro de Estudios Cient\'{\i}ficos (CECS), Casilla
1469, Valdivia, Chile}.} }
\date{}
\maketitle

\begin{abstract}
Some dynamical aspects of five-dimensional supergravity as a Chern-Simons
theory for the $SU(2,2|N)$ group, are analyzed. The gravitational sector is
described by the Einstein-Hilbert action with negative cosmological constant
and a Gauss-Bonnet term with a fixed coupling. The interaction between matter
and gravity is characterized by intricate couplings which give rise to
dynamical features not present in standard theories. Depending on the location
in phase space, the dynamics can possess different number of propagating
degrees of freedom, including purely topological sectors. This inhomogeneity of
phase space requires special care in the analysis.

Background solutions in the canonical sectors, which have regular dynamics with
maximal number of degrees of freedom, are shown to exist. Within this class,
explicit solutions given by locally AdS spacetimes with nontrivial gauge fields
are constructed, and BPS states are identified. It is shown that the charge
algebra acquires a central extension due to the presence of the matter fields.
The Bogomol'nyi bound for these charges is discussed. Special attention is
devoted to the $N=4$ case since then the gauge group has a $U(1)$ central
charge and the phase space possesses additional irregular sectors.

\end{abstract}


\section{Introduction}

Standard supergravity with a negative cosmological constant is a gauge theory
with fiber bundle structure only in three dimensions. Its Lagrangian is
described by a Chern-Simons (\textbf{CS}) form for the super-AdS group
$Osp(p|2)\otimes Osp(q|2)$ \cite{Achucarro-Townsend}. AdS supergravity theories
sharing this powerful geometrical structure can also be formulated in five
\cite{Chamseddine} and higher odd dimensions \cite {Troncoso-Zanelli}. These
theories are constructed assuming that the dynamical fields belong to a single
connection for a supersymmetric extension of the AdS group, and consequently,
the supersymmetry algebra closes automatically \emph{off-shell} without
requiring auxiliary fields \cite{Troncoso-Zanelli(off-shell)}. The existence of
an eleven-dimensional AdS supergravity theory which is gauge theory for
$OSp(32|1)$ exhibiting the features mentioned above opens up a number of new
questions, and is particularly interesting due to its possible connection with
M-theory \cite {Troncoso-Zanelli}. This problem has been further explored in
Refs. \cite{Horava}-\cite{deAzcarraga2}.

This elegant geometrical setting with its appealing gauge invariance leads,
however, to a rich and quite complex dynamics involving unexpected problems. In
order to understand better the subtleties, it is instructive to analyze the
simplest nontrivial CS system in some detail, which is the five dimensional
case. For the purely gravitational sector, the Lagrangian in $D=5$ dimensions
contains the Gauss-Bonnet term which is quadratic in the curvature, while for
$D\geq 7$, additional terms with higher powers of the curvature and explicitly
involving torsion are also required \cite{HDG}. The higher powers of curvature
give rise to interesting dynamical sectors within these theories which, even at
the linearized level, are beyond the notions learned from standard
supergravity.

In five dimensions, the locally supersymmetric extension of gravity with
negative cosmological constant was found in \cite{Chamseddine}, and generalized
in \cite{Troncoso-Zanelli} for higher odd dimensions. For vanishing
cosmological constant supergravity theories sharing this geometric structure
have also been constructed in
\cite{Hassaine-Troncoso-Zanelli,Banados-Troncoso-Zanelli,Hassaine-Olea-Troncoso}.

CS theories for $D\geq 5$ are not necessarily topological but contain
propagating degrees of freedom \cite{Banados-Garay-Henneaux}. Their dynamical
structure changes throughout phase space, changing drastically from purely
topological sectors to others with a large number of local degrees of freedom.
Sectors where the number of degrees of freedom is less than maximal are called
\emph{degenerate} and on them additional local symmetries emerge
\cite{Saavedra-Troncoso-Zanelli}.

Another unusual feature of these systems is that the symmetry generators (first
class constraints) may become functionally dependent in some regions of phase
space, called \emph{irregular} sectors. Dirac's canonical formalism cannot be
directly applied in these sectors, obscuring the dynamical content of CS
theories \cite{Henneaux-Teitelboim,Miskovic-Zanelli,Miskovic-Troncoso-Zanelli}.
These irregularities also imply that the theory is not correctly described by
its linearized approximation and hence the perturbative analysis cannot be
trusted \cite{Chandia-Troncoso-Zanelli,Miskovic,Miskovic-Pons}, the canonical
analysis breaks down and it is not clear how to identify the physical
observables (propagating degrees of freedom, conserved charges, etc.).

Degeneracy and irregularity are independent features that occur in any CS
theory for $D\geq5$ but are rarely found in field theories. They arise
naturally in fluid dynamics, as in the Burgers equation \cite{Burgers}, or in
the propagation of shock waves in compressible fluids described by the
Chaplygin and Tricomi equations \cite{Landau-Lifschitz}. Irregular sectors have
also been found in the Plebanski theory \cite{Buffenoir-Henneaux-Noui-Roche}.

Fortunately, the troublesome configurations generically occur in sets of
measure zero in phase space and one can always restrict the attention to open
sets where the canonical analysis holds. Such \emph{canonical} configurations
fill most of the phase space and it is desirable to know whether among them one
can find states that could be regarded as vacua around which a perturbatively
stable field theory can be built.

The presence of unbroken supersymmetries in backgrounds admitting Killing
spinors implies lower bounds for the sum of charges through the Bogomol'nyi
formula. This leads to the positivity of energy in standard supergravity
\cite{Abbott-Deser,Witten(E),Deser-Teitelboim,Grisaru(E)}, which also ensures
the stability of the configurations that saturate the energy bound (BPS
states). These configurations correspond to good perturbative vacua and in this
work it is shown that it is indeed possible to identify canonical
configurations which are BPS states.

In the next section, the Lagrangian of five-dimensional supergravity as a CS
theory for the supersymmetric extension of AdS$_5$, $SU(2,2|N)$, is reviewed.
Special attention is devoted to the case $N=4$ in which the gauge group
acquires a $U(1)$ central extension and the phase space possesses additional
irregular sectors. In Sect.3, the canonical representation of the charge
algebra, including its central extension, is constructed in a canonical sector
previously discussed in \cite{Miskovic-Troncoso-Zanelli}. In Sect.4 the
conditions on the background manifold that allow the existence of globally
defined Killing spinors are presented. The Killing spinors are explicitly given
in the canonical background and for a spatial boundary with topology
$S^1\otimes S^1\otimes S^1$. In Sect.5 the Bogomol'nyi bound is established,
and the conclusions and discussion are contained in Sect.6.


\section{AdS$_{5}$ supergravity as a Chern-Simons theory}

The supersymmetric extension of the AdS group in five dimensions is $SU(2,2|N)$
\cite{Nahm,Strathdee}, generated by the set $\mathbf{G}_{K}=
\{\mathbf{G}_{\bar{K}}, \mathbf{Z}\}$, where $\mathbf{Z}$ is the generator of
the $U(1)$ subgroup, and $\mathbf{G}_{\bar{K}} \equiv
\{\mathbf{J}_{ab},\mathbf{J}_{a};\mathbf{Q}_{s}^{\alpha },
\mathbf{\bar{Q}}_{\alpha }^{s};\mathbf{T}_{\Lambda}\}$. Here, $\mathbf{J}_{ab}$
and $\mathbf{J}_{a}$ are the generators of the AdS group $SO(4,2)$, and
$\mathbf{T}_{\Lambda }$ generate the $R$-symmetry group
$SU(N)$.\footnote{Hereafter, $a$,$b=0,\ldots ,4$ and $\alpha =1,\ldots ,4$
stand for vector and spinor indices in tangent space, respectively. The index
$s=1,\ldots ,N$ corresponds to a vector representation of $SU(N)$, whose
generators are labelled by $\Lambda =1,\ldots ,N^{2}-1$.} The supersymmetry
generators are given by $\mathbf{Q}_{s}^{\alpha }$ and
$\mathbf{\bar{Q}}_{\alpha }^{s}$, which transform as Dirac spinors in a vector
representation of $SU(N)$, and carry $U(1)$ charges $q=\pm \left(
\frac{1}{4}-\frac{1}{N}\right) $. The dimension of the superalgebra $su(2,2|N)$
is $\Delta =N^{2}+8N+15$. Its explicit form and a $(4+N)\times (4+N)$ matrix
representation for its generators are given in the Appendix.

Chern-Simons AdS$_{5}$ supergravity \cite{Chamseddine} is a gauge theory for
the Lie-algebra-valued connection 1-form $\mathbf{A}=A_{\mu }^{K}\mathbf{G}%
_{K}dx^{\mu }$, with components
\begin{equation}
\mathbf{A}=\frac{1}{\ell}\,e^{a}\mathbf{J}_{a}+\frac{1}{2}\,\omega ^{ab}
\mathbf{J}_{ab}+a^{\Lambda }\mathbf{T}_{\Lambda }+\left( \bar{\psi}_{\alpha
}^{s}\mathbf{Q}_{s}^{\alpha }-\mathbf{\bar{Q}}_{\alpha }^{s}\psi _{s}^{\alpha
}\right) +b\,\mathbf{Z}\;.  \label{connection}
\end{equation}
The bosonic sector of the theory contains the vielbein and the spin connection
$(e^{a},\omega ^{ab})$, the $SU(N)$ gauge field $a^{\Lambda }$ and the $U(1)$
field $b$. The fermionic fields $\psi _{s}$ are $N$ complex gravitini in a
vector representation of $SU(N)$.

The Lagrangian $L(\mathbf{A})$ satisfies
\begin{equation}
dL=k\,\left\langle \mathbf{F}^{3}\right\rangle =k\,g_{KLM}\,F^{K}F^{L}F^{M},
\label{dL 5}
\end{equation}
where $\mathbf{F}=d\mathbf{A}+\mathbf{A}^{2}=F^{K}\mathbf{G}_{K}$ is the
field-strength 2-form, and $k$ is a dimensionless constant.\footnote{Here we
omit the wedge symbol between forms for simplicity.} The bracket $\left\langle
\cdots \right\rangle $ stands for the supertrace in a representation which
naturally defines the invariant tensor $g_{KLM}$ which is (anti)symmetric under
permutation of (fermionic) bosonic indices \cite{Chandia-Troncoso-Zanelli} (see
Appendix). The action and its corresponding field equations are given by
\begin{equation}
I\left[ \mathbf{A}\right] =\int L(\mathbf{A})=k\int \left\langle
\mathbf{AF}^{2}-\frac{1}{2}\,\mathbf{A}^{3}\mathbf{F}+\frac{1}{10}\,
\mathbf{A}^{5}\right\rangle \;,  \label{I5}
\end{equation}
\begin{equation}
\left\langle \mathbf{F}^{2}\mathbf{G}_{K}\right\rangle =0\;. \label{Field Eqns
Bracket}
\end{equation}
The components of the field-strength $\mathbf{F}$ read
\begin{equation}
\mathbf{F}=F^{a}\mathbf{J}_{a}+\frac{1}{2}\,F^{ab}\mathbf{J}_{ab}+F^{\Lambda
}\mathbf{T}_{\Lambda }+\left( \nabla
\bar{\psi}^s\mathbf{Q}_s-\mathbf{\bar{Q}}^s\nabla \psi_s\right)
+F^{z}\,\mathbf{Z}\;, \label{F-components}
\end{equation} where
\begin{equation}
\begin{tabular}{llllll}
$F^{a}$ & $=$ & $\frac{1}{\ell }\,T^{a}+\frac{1}{2}\,\bar{\psi}^{s}\Gamma
^{a}\psi _{s}\,,$ & $F^{\Lambda }$ & $=$ & $\mathcal{F}^{\Lambda }
+\bar{\psi}^{s}\left( \tau ^{\Lambda }\right) _{s}^{r}\psi _{r}\,,\smallskip $ \\
$F^{ab}$ & $=$ & $R^{ab}+\frac{1}{\ell ^{2}}\,e^{a}e^{b}-\frac{1}{2}
\bar{\psi}^{s}\Gamma ^{ab}\psi _{s}\,,\qquad \qquad $ & $F^{z}$ & $=$ &
$f-i\bar{\psi}^{s}\psi _{s}\,.$
\end{tabular}
\end{equation}
Here the curvature and torsion two-forms have the form
\begin{equation}
R^{ab}=d\omega ^{ab}+\omega _{\;\,c}^{a}\omega ^{cb}\,,\qquad \qquad
\,T^{a}=de^{a}+\omega _{\;\,b}^{a}e^{b}\;,
\end{equation}
respectively, $f=db$ is the $u(1)$ field-strength, and $a\equiv a^{\Lambda
}\tau _{\Lambda }$, $\mathcal{F}=da+a^{2}$ are the connection and curvature for
$SU(N)$, where the $N\times N$ matrices $\tau ^{\Lambda }$ stand for the
$su(N)$ generators.

The components of the field-strength along the fermionic generators are given
by the AdS$_5\times$ $SU(N)$ $\times$ $U(1)$ covariant derivative\footnote{The
covariant derivative acts on a Lie-algebra valued $p$-form $\mathbf{X}_{p}$ as
$\nabla \mathbf{X}_{p}=\mathbf{X}_{p}+\left[
 \mathbf{A,X}_{p}\right] $, where $\left[ \mathbf{A,X}_{p}\right]
 =\mathbf{AX}_{p}-\left(
-\right) ^{p}\mathbf{X}_{p}\mathbf{A}$.}
\begin{equation}
\nabla \psi _{s}\equiv \left( D+\frac{1}{2\ell }\,e^{a}\Gamma _{a}\right) \psi
_{s}-a_{\;s}^{r}\psi _{r}+i\left( \frac{1}{4}-\frac{1}{N}\right) b\psi_{s},
\label{D psi}
\end{equation}
where $D\psi _{s}=\left( d+\frac{1}{4}\,\omega ^{ab}\Gamma _{ab}\right) \psi
_{s}$ is the Lorentz covariant derivative, and $\ell \,$is the AdS\ radius.

The decomposition (\ref{F-components}) allows to write the Lagrangian in a
manifestly Lorentz covariant way as
\begin{equation}
L=L_{G}\left( \omega ,e\right) +L_{SU(N)}\left( a\right) +L_{U(1)}\left( \omega
,e,b\right) +L_{F} \left( \omega ,e,a,b,\psi \right) \,,
\end{equation}
up to a boundary term. The gravitational sector is described by
\begin{equation}
L_{G} =  \frac{k}{8}\,\varepsilon _{abcde}\,\left(
\frac{1}{\ell}\,R^{ab}R^{cd}e^{e}+ \frac{2}{3\ell ^{3}}\,R^{ab}e^{c}e^{d}e^{e}+
\frac{1}{5\ell^{5}}\,e^{a}e^{b}e^{c}e^{d}e^{e}\right) \;,
\end{equation}
which is a linear combination of the Einstein-Hilbert Lagrangian with negative
cosmological constant and the Gauss-Bonnet term which is quadratic in the
curvature with a fixed coupling. The matter sector is described by
\begin{equation}
\begin{array}{lll}
L_{SU(N)} & = & ik\,\mbox{Tr}\left( \,a\mathcal{F}^{2}-
\frac{1}{2}\,a^{3}\mathcal{F}+\frac{1}{10}\,a^{5}\right)\,, \medskip \\
L_{U(1)} & = & -k\,\left( \frac{1}{4^{2}}- \frac{1}{N^{2}}\right) b\left(
db\right) ^{2}+\frac{3k}{4\ell ^{2}}\, \left( T^{a}T_{a}-\frac{\ell
^{2}}{2}\,R^{ab}R_{ab}-R^{ab}e_{a}e_{b}\right)
b-\frac{3k}{N}\,\mathcal{F}^{\Lambda }\mathcal{F}_{\Lambda }b\,,\medskip \\
L_{F} & = & -\frac{3ik}{4}\,\bar{\psi}^{s}\left[ \frac{1}{\ell } \,T^{a}\Gamma
_{a}+\frac{1}{2}\,\left( R^{ab}+\frac{1}{\ell^2}\,e^{a}e^{b}\right) \Gamma
_{ab}+2i\left( \frac{1}{N}+\frac{1}{4}\right) \,db-\bar{\psi}^{r}\psi
_{r}\right] \nabla \psi _{s}\medskip \\
&  & -\frac{3ik}{2}\bar{\psi}^{s}\left( \mathcal{F}_{s}^{r}-\frac{1}{2}\,
\bar{\psi}^{r}\psi _{s}\right) \nabla \psi _{r}+\mbox{c.c.\thinspace ,}
\end{array}
\end{equation}
where $\mathcal{F}_{r}^{s}=\mathcal{F}^{\Lambda }\left( \tau _{\Lambda }\right)
_{r}^{s}$.

Note that the case $N=4$ is exceptional and deserves special attention. As it
can be seen from the covariant derivative (\ref{D psi}) and the Lagrangian
$L_{U(1)}$, in this case gravitini become neutral under $U(1)$, and the
dynamics of the $U(1)$ field $b$ changes because it looses the cubic kinetic
term (the component $g_{zzz}$ of the invariant tensor vanishes). This reflects
the fact that for $N=4$, the $U(1)$ generator becomes a central charge in the
superalgebra $su(2,2|4)$ (see Appendix).

By construction, the action is invariant under diffeomorphisms and under
infinitesimal gauge transformations\emph{\ }$\delta _{\lambda
}\mathbf{A}=-\nabla \mathbf{\lambda }$, where $\mathbf{\lambda}$ is a Lie
algebra valued zero form. Local supersymmetry transformations can be obtained
as a particular case choosing the parameter as $\mathbf{\lambda }=
\bar{\epsilon}^{s}\mathbf{Q}_{s}-\mathbf{\bar{Q}}^{s}\epsilon _{s}$, from which
one obtains
\begin{equation}
\begin{array}{llllll}
\delta _{\epsilon }e^{a} & = & -\frac{1}{2}\,\left( \bar{\psi}^{s}\Gamma
^{a}\epsilon _{s}-\bar{\epsilon}^{s}\Gamma ^{a}\psi _{s}\right) \,,\qquad
\qquad \medskip & \delta _{\epsilon }\psi _{s} & = & -\nabla \epsilon _{s}\,,
\\
\delta _{\epsilon }\omega ^{ab} & = & \frac{1}{4}\,\left( \bar{\psi}^{s}\Gamma
^{ab}\epsilon _{s}-\bar{\epsilon}^{s}\Gamma ^{ab}\psi _{s}\right) \,,\medskip &
\delta _{\epsilon }\bar{\psi}^{s} & = & -\nabla \bar{\epsilon}
^{s}\,, \\
\delta _{\epsilon }a^{\Lambda } & = & \bar{\psi}^{s}\left( \tau ^{\Lambda
}\right) _{s}^{r}\epsilon _{r}-\bar{\epsilon}^{s}\left( \tau ^{\Lambda }\right)
_{s}^{r}\psi _{r}\,, & \delta _{\epsilon }b & = & i\left( \bar{\psi}
^{s}\epsilon _{s}-\bar{\epsilon}^{s}\psi _{s}\right) \,.
\end{array}
\label{SUSY}
\end{equation}
Note that, as a consequence, the supersymmetry algebra closes \emph{off-shell}
by construction, without requiring auxilliary fields
\cite{Troncoso-Zanelli(off-shell)}.


\section{Charges and their algebra in the canonical sectors}

In order to have a bona fide BPS bound, a canonical realization of the
supersymmetry algebra is needed. We follow the time-honored formalism of Dirac
for constrained systems since it ensures by construction the closure of the
canonical generators algebra. However, the standard Dirac procedure, required
to identify the physical observables (propagating degrees of freedom, conserved
charges, etc.), is not directly applicable around irregular backgrounds.
Indeed, the naive linearization of the theory fails to provide a good
approximation to the full theory around those backgrounds
\cite{Miskovic-Zanelli,Miskovic-Troncoso-Zanelli,Chandia-Troncoso-Zanelli}.

Thus, we analyze the system around background solutions in the canonical
sectors, namely, sectors possessing maximal number of degrees of freedom where
all constraints are functionally independent. The action (\ref{I5}) can be seen
to belong to the class of theories studied in \cite{Miskovic-Troncoso-Zanelli},
for which a family of backgrounds in the canonical sectors were identified. It
is worth mentioning that for $N=4$ the theory contains additional irregular
sectors which do not exist otherwise, and which require special attention.

As shown in \cite{Miskovic-Troncoso-Zanelli}, configurations where the only
nonvanishing components of $F^{\bar{K}}$ is\footnote{The five-dimensional
manifold is assumed to be topologically $\mathbb{R}\otimes \Sigma $, and the
coordinates are chosen as $x^{\mu }=(x^{0},x^{i})$, where $x^{i}$, with
$i=1,\ldots ,4$ correspond to the space-like section $\Sigma $.}
\begin{equation} \label{canonical1}
F_{12}^{\bar{K}}\,dx^{1}dx^{2}\neq 0\,,
\end{equation}
for at least one $\bar{K}$ and
\begin{equation} \label{canonical2}
F_{34}^{z}=0\,,\qquad \mbox{with  }\qquad \det \left( F_{ij}^{z}\right) \neq
0\, ,
\end{equation}
turn out to be canonical for any $N$. Therefore, around this kind of background
solutions the counting of degrees of freedom can be safely done following the
standard procedure \cite{Henneaux-Teitelboim,Henneaux}. In this case, the
number is $\Delta-2 =N^2+8N+13$ (see \cite{Banados-Garay-Henneaux}).

\subsection{Charge algebra}

The advantage of the class of canonical sectors described above, is that the
splitting between first and second class constraints, which is in general an
extremely difficult task, can be performed explicitly. As a consequence, the
conserved charges and their algebra can be obtained following the
Regge-Teitelboim approach \cite{Regge-Teitelboim}, and as shown in
\cite{Miskovic-Troncoso-Zanelli}, they turn out to be
\begin{equation}
Q\left[ \lambda \right] =-3k\int\limits_{\partial \Sigma }g_{KLM}\,\lambda
^{K}\bar{F}^{L}A^{M}.  \label{charge}
\end{equation}
Here $\bar{F}$ is the background field strength and the parameters $\lambda
^{K}(x)$ approach covariantly constant fields at the boundary. According to the
Brown-Henneaux theorem, in general the charge algebra is a central extension of
the gauge algebra \cite{Brown-HenneauxQ},
\begin{equation}
\left\{ Q\left[ \lambda \right] ,Q\left[ \eta \right] \right\} =Q\left[ \left[
\lambda ,\eta \right] \right] +C\left[ \lambda ,\eta \right] \,. \label{charge
algebra}
\end{equation}
In the present case the central charge is
\begin{equation}
C\left[ \lambda ,\eta \right] =3k\int\limits_{\partial \Sigma }g_{KLM}\,\lambda
^{K}\bar{F}^{L}d\eta ^{M}.  \label{central charge}
\end{equation}

The charge algebra can be recognized as the WZW$_{4}$ extension of the full
gauge group \cite{Losev-Moore-Nekrasov-Shatashvili}. In an irregular sector the
charges are not well defined and the naive application of the Dirac formalism
would at best lead to a charge algebra associated to a subgroup of $G$.

Having obtained the canonical realization of the symmetry algebra, allows one
to proceed with construction of the BPS bound as well as the states that
saturate it. In the next section we find explicit BPS solutions within the
class of canonical configurations given by Eqs. (\ref{canonical1}) and
(\ref{canonical2}), and in Sect.5, we explicitly obtain the Bogomol'nyi bound
for states in the neighborhood of a BPS state.


\section{Background solutions}

The simplest background solutions are purely bosonic ($\psi _{s}=0$), for which
the field equations become
\begin{eqnarray}
\varepsilon _{abcde}\left( R^{bc}R^{de}+\frac{2}{\ell ^{2}}
\,R^{bc}e^{d}e^{e}+\frac{1}{\ell ^{4}}\,e^{b}e^{c}e^{d}e^{e}\right)
+\frac{4}{\ell }\,T_{a}\,f &=&0\,,
\label{EGB-Eq} \\
\varepsilon _{abcde}\left( R^{cd}+\frac{1}{\ell ^{2}}\,e^{c}e^{d}\right)
T^{e}+\ell \left( R_{ab}+\frac{1}{\ell ^{2}}\,e_{a}e_{b}\right) \,f &=&0\,,
\label{Torsion-Eq} \\
\frac{1}{4\ell ^{2}}\,\left( \frac{\ell ^{2}}{2}R^{ab}R_{ab}+
\,R^{ab}e_{a}e_{b}-T^{a}T_{a}\right) +\frac{1}{N}\,\mathcal{F}^{\Lambda }
\mathcal{F}_{\Lambda }-\left( \frac{1}{N^{2}}-\frac{1}{4^{2}}\right) \,f\,f
&=&0\,,  \label{U(1)-Eq} \\
\gamma _{\Lambda \Lambda _{1}\Lambda _{2}}\mathcal{F}^{\Lambda
_{1}}\mathcal{F}^{\Lambda _{2}}+\frac{2}{N}\,\mathcal{F}_{\Lambda }\,f &=&0\,.
\label{SU(N)-Eq}
\end{eqnarray}
Assuming spacetime to be locally AdS, so that
$F^{ab}=R^{ab}+\frac{1}{\ell^{2}}\,e^{a}e^{b}=0$, the torsion vanishes.
Therefore, the modified Einstein and torsion Eqs. (\ref{EGB-Eq}, \ref
{Torsion-Eq}) are trivially satisfied, and the first term in Eq. (\ref
{U(1)-Eq}) vanishes, as well.

Note that in the absence of matter fields, any locally AdS spacetime solves the
bosonic fields equations. However, this kind of backgrounds are maximally
degenerate and irregular. It is noteworthy that in this case it is possible to
overcome degeneracy and irregularity by switching on matter fields which do not
have a back reaction on the metric.\footnote{Matter fields may not produce back
reaction as a result of non-minimal couplings. This has been observed in very
simple systems, such as general relativity with scalar fields
\cite{Ayon-Beato:2005tu,Robinson:2006ib}.}

It must be emphasized that locally AdS spacetime configurations require the
presence of nontrivial $SU(N)$ and $U(1)$ fields. Indeed, it might seem as if a
simpler solution could be obtained for $N=4$ by turning off the $SU(4)$
curvature, ${\cal F}^{\Lambda }=0$ in Eqs. (\ref{Backgrounds})
\cite{Chandia-Troncoso-Zanelli}. That solution is, however, irregular.

As required by (\ref{canonical1}), locally AdS spacetime configurations must
have the $SU(N)$ field-strength ${\cal F}_{12}^{\Lambda}$ switched on.  It is
easy to see that this configuration solves the remaining field equations
(\ref{U(1)-Eq}) and (\ref{SU(N)-Eq}), provided the $U(1)$ field $b$ has a
field-strength satisfying $F_{34}^{z}=0$, while the remaining components are
arbitrary, and $F_{ij}^{z}=\partial _{i}b_{j}-\partial _{j}b_{i}$ can be
assumed to be invertible.

In sum, the bosonic solutions given by
\begin{eqnarray}
R^{ab} &=&-\frac{1}{\ell^{2}}\,e^{a}e^{b}\;,  \nonumber \\
T^{a} &=&0\;,  \nonumber \\
F^{\Lambda } &=&{\cal F}_{12}^{\Lambda }\,dx^{1}dx^{2}\neq 0\;,
\label{Backgrounds}
\\
F_{34}^{z} &=&0\,, \qquad\qquad \mbox{with\ }\det \left( F_{ij}^{z}\right) \neq
0\;, \nonumber
\end{eqnarray}
provide canonical backgrounds for any $N$.

One then concludes that in this supergravity theory, constant curvature
spacetimes can be embedded in a canonical sector since they can be consistently
combed with nontrivial $SU(N)$ and $U(1)$ fields. This includes AdS spacetime
and quotients of it, as in Refs.\cite{AdS-Quotients}, giving rise to a wide
class of solutions with different topologies.

In what follows we will look for solutions of the form (\ref{Backgrounds})
admitting Killing spinors.

\subsection{BPS states}

Bosonic solutions of the field equations which are left invariant under
globally defined supersymmetry transformations (BPS states), by virtue of Eqs.
(\ref {SUSY}) must satisfy $\delta _{\epsilon }\psi _{s}=-\nabla \epsilon
_{s}=0$. Hence, Killing spinors $\epsilon _{s}$ solve the equation

\begin{equation}
\nabla \epsilon _{s}=\left( d+\textbf{A}_{AdS}\mathbf{+}i\left( \frac{1}{4}-
\frac{1}{N}\right) b\right) \epsilon _{s}-a_{\;s}^{r}\epsilon _{r}=0\;,
\label{Killing Spinor Eq}
\end{equation}
where $a_{\;r}^{s}=a^{\Lambda }(\tau _{\Lambda })_{\;r}^{s}$, and the AdS
connection is given by $\textbf{A}_{AdS}=\frac{1}{4}\,\omega ^{ab}\Gamma
_{ab}+\frac{1}{2\ell} \,e^{a}\Gamma_{a}$. The consistency condition of the
Killing spinor Eq. (\ref {Killing Spinor Eq}), $\nabla \nabla \epsilon _{s}=0$,
reads
\begin{equation}
\left( \textbf{F}_{AdS}\mathbf{+}i\left( \frac{1}{4}-\frac{1}{N}\right)
f\right) \epsilon _{s}-{\cal F}_{\;s}^{r}\epsilon _{r}=0\;,  \label{Consistency
KS}
\end{equation}
where the AdS curvature is
\[
\textbf{F}_{AdS}=\frac{1}{4}\,\left(
R^{ab}+\frac{1}{\ell^{2}}\,e^{a}e^{b}\right) \Gamma
_{ab}+\frac{1}{2\ell^{2}}\,T^{a}\Gamma _{a}\;.
\]

Note that for $N=4$, neither the Killing spinor equation (\ref{Killing Spinor
Eq}) nor the consistency condition (\ref{Consistency KS}) involve the $U(1)$
field. Hence, for simplicity, we will focus on this case in what follows.

\subsubsection{$N=4$}

For $N=4$, equation (\ref{Killing Spinor Eq}) reduces to
\begin{equation}
\left[ \left( d+\textbf{A}_{AdS}\right) \delta _{s}^{r}-a_{s}^{r}\right]
\epsilon _{r}=0\,,  \label{Killing Eq N=4}
\end{equation}
and since the AdS\ curvature $\textbf{F}_{AdS}$ vanishes for the class of
background solutions under consideration given by (\ref{Backgrounds}), the
consistency condition simply reads
\begin{equation}
{\cal F}_{s}^{r}\epsilon _{r}=0\,.  \label{Consistency N=4}
\end{equation}

For the canonical class of solutions given by (\ref{Backgrounds}), the
consistency condition (\ref{Consistency N=4}) means that the Killing spinors
must be zero modes of the $SU(4)$ field strenght. Hence, $\mathcal{F} ^{\Lambda
}$ must be nonvanishing for more than one value of the index $ \Lambda $, so
that the contributions of all components cancel.

As an example, taking advantage of the isomorphism $su(4)\simeq so(6)$, the
$SU(4)$ curvature can be expressed as $\mathcal{F}_{r}^{s}=\frac{1
}{2}\mathcal{F}^{IJ}\left( \tau _{IJ}\right) _{r}^{s}$, where the $so(6)$
generators
\begin{equation}
\tau _{IJ}=\frac{1}{2}\,\hat{\Gamma}_{IJ}\,,\qquad \left( I,J=1,\ldots
,6\right)
\end{equation}
are given in terms of the Euclidean Dirac matrices $\hat{\Gamma}_{I}$, with
$\hat{\Gamma}_{IJ}=\frac{1}{2}\,\left[ \hat{\Gamma}_{I},\hat{\Gamma}_{J}\right]
$. The commuting matrices $\tau _{12}$ and $\tau _{34}$ generate a $U(1)\otimes
U(1)$ subgroup for $SU(4)$, and since $\left( \tau _{12}\right) ^{2}=\left(
\tau _{34}\right) ^{2}=-\frac{1}{4}$, the eigenvalues of $\tau _{12}$ and $\tau
_{34}$ are $\pm \frac{i}{2}$.

For simplicity, one can make use of a ``twisted'' configuration for which the
only nonvanishing $U(1)\otimes U(1)$ components of the $SU(4)$ curvature are
given by $\mathcal{F}^{12}=da^{12}$ and $\mathcal{F}^{34}=da^{34}$, and the
Killing spinor $\epsilon_s$ is assumed to satisfy
\begin{equation}
\left( \tau _{12}\right)_s^r\epsilon _r =\frac{i}{2}\,\epsilon_s\,,\qquad
\qquad \left( \tau _{34}\right)_s^r\epsilon _r =-\frac{i}{2}\,\epsilon_s\;.
\label{twist}
\end{equation}
Therefore, the consistency condition (\ref{Consistency N=4}) becomes
\[
\frac{i}{2}\,\left( \mathcal{F}^{12}-\mathcal{F}^{34}\right) \, \epsilon_s=0\;,
\]
and is solved by $\mathcal{F}^{12}=\mathcal{F}^{34}$. Then the $SU(4)$ field
satisfy
\begin{equation}
a^{12}=a^{34}+d\theta \;,  \label{1234}
\end{equation}
where $\theta =\theta \left( x\right) $ is an arbitrary phase. Hence, the
twisted $SU(4)$ configuration satisfies $a_{r}^{s}\,\epsilon _{s}=\frac{i}{2}
\,d\theta \,\epsilon _{r}$, and the Killing equation (\ref{Killing Eq N=4})
reduces to
\begin{equation}
\left( d+\textbf{A}_{AdS}-\frac{i}{2}\,d\theta \right) \,\epsilon _s=0\;.
\label{Killing Eq Twisted }
\end{equation}
The solution of this last equation is given by
\begin{equation} \label{epsilon}
\epsilon_s=e^{\frac{i}{2}\,\theta }\eta _s\;,
\end{equation}
where $\eta _s$ satisfies the twisting conditions (\ref {twist}), and is a
nontrivial solution of the Killing spinor equation in vacuum
\begin{equation} \label{ADSspinor}
\left( d+\textbf{A}_{AdS} \right) \,\eta _s=0\;.
\end{equation}
Locally AdS spacetimes admitting Killing spinors have been extensively
discussed in the literature \cite{AdS-Quotients}. We now consider a particular
geometry which is simple and allows to deal with a nontrivial topology at the
boundary.

\subsubsection{Explicit BPS solutions in the canonical sectors}

The local coordinates on $\mathcal{M}$ are chosen as $x^{\mu }=\left( t,\rho
,\varphi ^{p}\right) $, where $\varphi ^{p}$ $\left( p=2,3,4\right) $
parametrize the boundary $\partial \Sigma ,$ placed at the infinity of the
coordinate $\rho $ $\left( \rho \geq 0\right) $.

The globally AdS space-time ($F^{AB}=0$) can be described by the metric
\begin{equation}
ds_{AdS}^{2}=\ell ^{2}\,\left( d\rho ^{2}+e^{2\rho }\eta _{\bar{p}
\bar{q}}\,dx^{\bar{p}}dx^{\bar{q}}\right) \,,  \label{AdS metric}
\end{equation}
where $x^{\bar{p}}=\left( t,\varphi ^{p}\right) $ and $\eta _{\bar{p}\bar{q}}=$
diag $\left( -,+,+,+,\right) $. The AdS connection is
\[
\mathbf{A}_{AdS}=\frac{1}{4}\,\omega ^{ab}\Gamma _{ab}+\frac{1}{2\ell }
\,e^{a}\Gamma _{a}=\frac{1}{2\ell }\,\left[ e^{1}\Gamma _{1}+e^{\bar{p}}\Gamma
_{\bar{p}}\left( 1+\Gamma _{1}\right) \right] \,.
\]
The Killing spinors for the metric (\ref{AdS metric}) solving equation
(\ref{ADSspinor}), have the form \cite{Lu-Pope-Townsend}
\begin{equation}
\eta_s =e^{-\frac{\rho }{2}\,\Gamma _{1}}\,\left[ 1-x^{\bar{p}} \Gamma
_{\bar{p}}\left( 1+\Gamma _{1}\right) \right] \,\eta_{0s}\,, \label{AdS
Killing}
\end{equation}
where $\eta_{0s}$ is a constant spinor. This gives a solution to the Killing
spinor equation (\ref{Killing Eq Twisted }) of the form (\ref{epsilon}),
provided $\eta_{0s}$ satisfies the twisting conditions (\ref{twist}).

Assuming the boundary of the spatial section to be topologically $\partial
\Sigma \simeq S^{1}\otimes S^{1}\otimes S^{1}$, the spinor $\eta_{s}$ must be
antichiral under the action of $\Gamma _{1}$ in order to be globally defined.
Then the solution becomes,
\begin{equation}
\epsilon _{s}=e^{\frac{1}{2}\,\left( i\theta +\rho\right) }\,\eta_{0s}\,,\qquad
\mbox{with} \qquad \left( 1+\Gamma _{1}\right) \,\eta_{0s}=0\,.
\end{equation}
Chirality and twisting conditions give $\frac{1}{8}\times 16=2$ unbroken
supersymmetries.

The remaining $SU(4)$ and $U(1)$ fields can be chosen as follows,
\begin{eqnarray}
\bar{a}^{12} &=&h\,d\rho \,,\label{a12}\\
\bar{a}^{34} &=&h\, d\rho \,-d\theta \,,\label{a34} \\
 \bar{b}&=&E\varphi^3 \,d\rho + B \varphi^4 d\varphi^2 \,,\label{b0}
\end{eqnarray}
where $h=h(\varphi^2)$ is an arbitrary function, and $E$, $B$ are nonvanishing
constants. Then $\bar{f}_{34}=0$, and $\det \left( \bar{f}_{ij}\right) = \left(
BE\right) ^{2}\neq 0$, as required by (\ref{Backgrounds}).

\subparagraph{Various topologies.}

For chosen boundary conditions $\mathbf{A}\rightarrow \mathbf{\bar{A}}$ and $
\mathbf{\lambda }\rightarrow \mathbf{\bar{\lambda}}$, where $\overline{\nabla}
 \mathbf{\bar{\lambda}}=0$, the most general central charge is given by
(\ref{central charge}). In locally AdS spacetimes, it
takes the form%
\begin{equation}
C\left[ \lambda ,\eta \right] =6k\int\limits_{\partial \Sigma }\left\langle
\mathbf{\lambda\, \bar{F}}_{U(1)}d\mathbf{\eta }\right\rangle
+6k\int\limits_{\partial \Sigma }\left\langle \mathbf{\lambda\,
\bar{F}}_{SU(4)}d\mathbf{\eta }\right\rangle \,,  \label{general C}
\end{equation}
so that the charge only acquires contributions from the internal bosonic
subgroup $SU(4)\otimes U(1)$.

If the topology of the boundary is isomorphic to $S^{1}\otimes S^{1}\otimes
S^{1}$, then the first term in (\ref{general C}) gives a non-trivial
contribution to the central charge since $\pi _{1}\left( U(1)\right) =
\mathbb{Z}$, while the second term vanishes because of $\pi _{1}\left(
SU(4)\right) =0$. When $SU(4)$ is explicitly broken into $U(1)\otimes
U(1)\otimes U(1)$, it can give non-trivial contribution, as well.

In the case of $S^{1}\otimes S^{2}$, since $\pi _{2}\left( G\right) =0$ for all
Lie groups, a non-trivial central charge can be obtained only if the Abelian
field $b$, or some other corresponding to a $U(1)$ subset of $SU(4)$, winds
around $S^{1}$.

If the topology is $S^{3}$, the non-trivial $SU(4)$ field must have three
non-vanishing components at $\partial \Sigma $, $\bar{F}_{pq}^{\Lambda }\neq
0$, where it is a solution if constraints. Then the first term in the central
charge vanishes due to $\pi _{3}\left( U(1)\right) =0$, but considering that
$\pi _{3}\left( SU(4)\right) =\mathbb{Z}$, the second term may give a
non-vanishing result.


\section{Bogomol'nyi bound}

In this section we establish the Bogomol'nyi bound \cite{Bogomolnyi} for
configurations in the neighborhood of a BPS state. Due to the presence of a
central charge in this case the bound cannot be obtained naively from the
original supersymmetry algebra.

Around the background (\ref{AdS metric}, \ref{a12}--\ref{b0}), the charges
(\ref{charge}) take the form\footnote{The normalization of the volume element
of $\partial \Sigma =S^{1}\otimes S^{1}\otimes S^{1}$ has been chosen as
$d^{3}x=\frac{d^{3}\varphi }{(2\pi )^{3}}$.}
\begin{equation}
Q\left[ \lambda \right] =\int \frac{d^{3}\varphi }{\left( 2\pi \right) ^{3}}
\,\lambda ^{K}q_{K}\,,\qquad q_{K}=\frac{3kB}{2}\,\gamma _{KL}\,A_{3}^{L}\,.
\label{Q evaluated}
\end{equation}
Note that the class of solutions considered here has identically vanishing
$u(1)$ charge since $\gamma _{zz}=\gamma _{\bar{K}z}=0$ (see Appendix). Thus,
the canonical algebra (\ref{charge algebra}) is a central extension of
$psu(2,2\left\vert 4\right.)$, and the central charge (\ref{central charge}) is
just
\begin{equation}
C\left[ \lambda ,\eta \right] =-\frac{3kB}{2}\,\int \frac{d^{3}\varphi }{
\left( 2\pi \right) ^{3}}\,\gamma _{KL}\,\lambda ^{K}\partial _{3}\eta ^{L}.
\label{C explicitly}
\end{equation}
In particular, $C\left[ \lambda ^{z},\eta ^{K}\right] \equiv 0$,\ which implies
that there is no $u(1)$ central extension.

Demanding the charges to vanish on the BPS background, we obtain \footnote{The
covariantly constant AdS vectors $\bar{\lambda}^{AB}$ are solutions of $ \nabla
_{AdS}\bar{\lambda}^{AB}=0$ with $\mathbf{A}_{AdS}$ given by (\ref{AdS
metric}). They have the form $\bar{\lambda}^{\bar{p}1}=\bar{\lambda}
^{\bar{p}5}=V^{\bar{p}}e^{\rho }$ ($V^{\bar{p}}=$ Const.). For the $su(4)$
connection (\ref{a12}, \ref{a34}), the nonvanishing covariantly constant
vectors, $\nabla_{su(4)}\bar{\lambda}^{IJ}=0,$ are
$\bar{\lambda}^{12},\bar{\lambda}^{34}, \bar{\lambda}^{56}=$ Const., and they
describe the unbroken symmetry $U(1)\otimes U(1)\otimes U(1)\subset SU(4)$ of
the background.}
\begin{eqnarray}
\bar{Q}_{u(1)} &=&Q\left[ \bar{\lambda}^{z},\bar{A}\right] =0\,,
\nonumber \\
\bar{Q}_{AdS} &=&Q\left[ \bar{\lambda}^{AB},\bar{A}\right] =\frac{3kB}{2}
\left( \bar{\lambda}^{31}-\bar{\lambda}^{35}\right) e^{\rho }=0\,, \\
\bar{Q}_{su(4)} &=&Q\left[ \bar{\lambda}^{IJ},\bar{A}\right] =
-\frac{3kB}{2}\,\bar{\lambda}^{34}\int \frac{d^{3}\varphi }{\left( 2\pi \right)
^{3}}\,
\partial _{3}\theta \left( \rho ,\varphi \right) \,.  \nonumber
\end{eqnarray}
Therefore, the phase $\theta $ is periodic in $\varphi^{3}$ in order to have a
vanishing $su(4)$ charge.

\subsection{Mode expansion}

Since the canonical BPS background has a boundary with topology $\partial
\Sigma \simeq S^{1}\otimes S^{1}\otimes S^{1}$, any variable $X\left( \rho
,\varphi \right) ,$ periodic or anti-periodic in $\vec{\varphi}=\left( \varphi
^{2},\varphi ^{3},\varphi ^{4}\right) \in \lbrack 0,2\pi )$, can be expanded in
Fourier series as
\begin{equation}
X\left( \rho ,\varphi \right) =\sum\limits_{\vec{\nu}}X_{\vec{\nu}}\left( \rho
\right) \,e^{i\,\vec{\nu}\cdot \vec{\varphi}},\qquad \qquad X_{\vec{\nu}
}\left( \rho \right) =\int \frac{d^{3}\varphi }{\left( 2\pi \right) ^{3}}
\,X\left( \rho ,\varphi \right) \,e^{-i\,\vec{\nu}\cdot \vec{\varphi}},
\end{equation}
where $\vec{\nu}\equiv \left( \nu _{2},\nu _{3},\nu _{4}\right) $ are winding
numbers. Bosonic modes are periodic in $\vec{\varphi}$ and the numbers $\vec{
\nu}$ are integers. Fermionic modes can be periodic (Ramond (\textbf{R})
sector), or anti-periodic (Neveu-Schwartz (\textbf{NS}) sector) in any of the
angular coordinates $\vec{\varphi}$, and the corresponding winding numbers are
integers ($\nu _{i}\in \mathbb{Z}$) or half-integers ($\nu _{i}+1/2\in
\mathbb{Z}$), respectively, giving rise to eight possible sectors
R$_{2}$-R$_{3}$-R$_{4}$, R$_{2}$-R$_{3}$-NS$_{4}$, etc.

The mode expansion of the charges (\ref{Q evaluated}) is
\begin{equation}
Q\left[ \lambda \right] =\sum\limits_{\vec{\nu}}\lambda _{\vec{\nu}
}^{K}q_{K,-\vec{\nu}}\,,\qquad q_{K,\vec{\nu}}=\frac{3kB}{2}\,\gamma
_{KL}\,A_{3,\vec{\nu}}^{L}\,,
\end{equation}
and the central charges (\ref{C explicitly}) are
\begin{equation}
C\left[ \lambda ,\eta \right] =-\frac{3ikB}{2}\,\gamma _{KL}\sum_{\vec{\nu},
\vec{\mu}}\mu _{3}\,\lambda _{\vec{\nu}}^{K}\,\eta _{\vec{\mu}}^{L}\,\delta
_{\vec{\nu}+\vec{\mu},\vec{0}}\,.
\end{equation}
Finally, the charge algebra acquires the form
\begin{equation}
\left\{ q_{K,\vec{\nu}},q_{L,\vec{\mu}}\right\} =f_{KL}^{\quad \;M}q_{M,\vec{
\nu}+\vec{\mu}}-\frac{3ikB}{2}\,\nu _{3}\,\gamma _{KL}\,\delta _{\vec{\nu}+
\vec{\mu},\vec{0}}\,.  \label{mode algebra}
\end{equation}
The algebra (\ref{mode algebra}) is a supersymmetric extension of the WZW$
_{4}$ algebra
\cite{Banados-Garay-Henneaux,Losev-Moore-Nekrasov-Shatashvili,Floreanini-Percacci-Rajaraman,Nair-Schiff,Gegenberg}.
It has a nontrivial central extension for $psu(2,2\left\vert 4\right. )$ which
depends only on $u(1)$ flux determined by $B$. Note that the modes
$q_{K,\vec{\nu}}$ with $\vec{\nu}=(0,\nu _{3},0)$ form a Kac-Moody subalgebra
with the central charge $c=-\frac{3kB}{2}$, while the modes with
$\vec{\nu}=(\nu _{2},0,0)$ and $(0,0,\nu _{4})$ form Kac-Moody subalgebras
without central charges.

\subsection{Bogomol'nyi bound}

In the case of fermionic charges, the algebra (\ref{mode algebra}) reads
\begin{equation}
\begin{array}{ll}
\left\{ q_{r,\vec{\nu}}^{\alpha },\bar{q}_{\beta ,\vec{\mu}}^{s}\right\}
=\medskip  & -\frac{1}{2}\,\delta _{r}^{s}\left( \Gamma ^{a}\right) _{\beta
}^{\alpha }\,q_{a,\vec{\nu}+\vec{\mu}}+\frac{1}{4}\,\delta _{r}^{s}\left(
\Gamma ^{ab}\right) _{\beta }^{\alpha }\,q_{ab,\vec{\nu}+\vec{\mu}} \\
& -\frac{1}{2}\,\delta _{\beta }^{\alpha }\,\left( \tau ^{IJ}\right)
_{r}^{s}\,q_{IJ,\vec{\nu}+\vec{\mu}}+\frac{3ikB}{2}\,\nu _{3}\,\delta
_{r}^{s}\,\delta _{\beta }^{\alpha }\,\delta _{\vec{\nu}+\vec{\mu},\vec{0} }\,.
\end{array}
\label{fermimodes}
\end{equation}
Multiplying (\ref{fermimodes}) by $\Gamma ^{0}$, and using the fact that the
operator $\left\{ q_{r,\vec{\nu}}^{\alpha },(q^{\dagger })_{\beta ,-\vec{\nu}
}^{s}\right\} $ is positive semidefinite, we have the bound
\begin{equation}
-\frac{1}{2}\,\delta _{r}^{s}\left( \Gamma ^{a}\Gamma ^{0}\right) _{\beta
}^{\alpha }\,q_{a,\vec{0}}+\frac{1}{4}\,\delta _{r}^{s}\left( \Gamma
^{ab}\Gamma ^{0}\right) _{\beta }^{\alpha }\,q_{ab,\vec{0}}-\frac{1}{2}
\,\left( \Gamma ^{0}\right) _{\beta }^{\alpha }\,\left( \tau ^{IJ}\right)
_{r}^{s}\,q_{IJ,\vec{0}}+\frac{3ikB}{2}\,\nu _{3}\,\left( \Gamma ^{0}\right)
_{\beta }^{\alpha }\delta _{r}^{s}\geq 0\,. \label{Bound}
\end{equation}
Decomposing the AdS boost charge as $q_{a,\vec{0}}=\left(
q_{0,\vec{0}},q_{\bar{a},\vec{0}}\right) $, one finds\footnote{ In this
signature, $\left( \Gamma ^{0}\right) ^{2}=-1$.}
\[
-\frac{1}{2}\,\delta _{r}^{s}\left( \Gamma ^{a}\Gamma ^{0}\right) _{\beta
}^{\alpha }\,q_{a,\vec{0}}=\frac{1}{2}\,\delta _{r}^{s}\delta _{\beta }^{\alpha
}\,E-\frac{1}{2}\,\delta _{r}^{s}\left( \Gamma ^{\bar{a}}\Gamma ^{0}\right)
_{\beta }^{\alpha }\,q_{\bar{a},\vec{0}}\,,
\]
where the energy is identified as $E=q_{0,\vec{0}}$. Then (\ref{Bound}) can be
rewritten as
\begin{equation}
\delta _{r}^{s}\delta _{\beta }^{\alpha }\,E\geq\delta _r^s\left( \Gamma
^{\bar{a}}\Gamma^0\right)_{\beta }^{\alpha }\,q_{\bar{a},\vec{0}}-
\frac{1}{2}\,\delta_r^s\left( \Gamma^{ab}\Gamma^0\right)_{\beta }^{\alpha
}\,q_{ab,\vec{0}}+\left( \Gamma ^0\right) _{\beta }^{\alpha }\,\left( \tau
^{IJ}\right) _r^s\,q_{IJ,\vec{0}}-3ikB\nu _{3}\,\left( \Gamma ^{0}\right)
_{\beta }^{\alpha }\delta _{r}^{s}\,. \label{inequality}
\end{equation}
The eigenvalues ($\lambda_i$) of the matrix $M^{s\alpha}_{r\beta}$ on the
r.h.s. of (\ref{inequality}) can be calculated from $\sum_i\lambda_i^2=$
Tr$(\mathbf{M})^2$, using the orthogonality of the group generators. The result
is\footnote{Apart from the traces of unit matrices, the other non-vanishing
traces are Tr$ (\Gamma^{a}\Gamma^{b})=4\eta ^{ab}$, Tr$(\Gamma ^{ab}\Gamma
^{cd})=-4\eta ^{\lbrack ab][cd]}$ and Tr$(\tau ^{IJ}\tau ^{I^{\prime }J^{\prime
}})=-\delta ^{\lbrack IJ][I^{\prime }J^{\prime }]}$.} $\lambda
^2=p^2+(3kB)^2\nu_3^2$, with $p^2\equiv \sum_{\bar{K}}(q_{\bar{K},\vec{0}})^2$.
The requirement  that the energy is not smaller than the largest eigenvalue of
$\mathbf{M}$, namely $ \delta _{r}^{s}\delta _{\beta }^{\alpha }\,E\geq
M^{s\alpha}_{r\beta}$, leads to the Bogomol'nyi bound
\begin{equation}
E\geq \sqrt{p^2+(3kB)^2(\nu _3)_{\min }^2}\,.
\end{equation}
This bound is saturated for the BPS states, $E_{BPS}=|3kB\nu_{3\,\min}|$\,. In
the NS$_{3}$ sector where $\left( \nu _{3}\right) _{\min }=\frac{1}{2}$, the
energy is $E_{BPS}= \left|\frac{3kB}{2}\right|$, while in the R$ _{3}$ sector,
$\left( \nu _{3}\right) _{\min }=0$, $E_{BPS}=0$. The ground state (\ref{AdS
metric}, \ref{a12}--\ref{b0}) is a R$_3$ state, consistent with the fact that
$\theta$ is periodic in $\varphi^3$.


\section{Conclusions}

Degeneracy and irregularity are largely unexplored phenomena in dynamical
systems. Although these features are rarely found in standard field theories,
they are unavoidable in higher dimensional gravity theories of current interest
such as those described by the Gauss-Bonnet and Lovelock actions.
Irregularities imply that the degrees of freedom of the linearized
approximation do not correspond to those of the full theory and hence one needs
to go beyond the perturbative analysis. Moreover, the canonical analysis breaks
down and since it is not clear how to identify the conserved charges --and
physical observables in general--, the possibility of finding a concrete
expression of their algebra is severely limited.

Although canonical sectors, which are nondegenerate and regular, fill open sets
of phase space, they may not be not easily identified. The rich geometric
structure of the supergravity theory considered here helps in this task, as
well as in the obtaining a canonical representation of the conserved charges.
It is found that, unlike the situation in standard theories, the resulting
charge algebra turns out to be a nontrivial central extension of the symmetry
algebra, in an analogous way as it occurs in the case of asymptotically AdS
gravity in three dimensions \cite{Brown-Henneaux}. In this case, the central
charge is nonzero thanks to the presence of matter fields with a nontrivial
winding. Interestingly, these matter fields have a nontrivial field strength
but nevertheless, due the nonminimal coupling, produce no back reaction on the
geometry.

The BPS bound is constructed here in the canonical sector. The canonical
realization of the algebra guarantees the stability of the theory, which would
not be achieved through the naive bound, constructed purely from the symmetry
algebra. BPS states in the canonical sectors saturating the bound are
explicitly found.

Conserved charges for Chern-Simons gravity theories in higher dimensions were
constructed using a background-independent approach and have been shown to be
well defined even for degenerate and irregular configurations, including black
holes \cite{MOTZ1}. These charges were shown to be related to the notion of
transgression forms \cite{MOTZ2}. Alternative expressions for conserved charges
also based on the idea of transgression forms have been constructed in Refs.
\cite{BFFF,IRS1,Raros,IRS2}. It would be interesting to see whether the
centrally extended algebra constructed here can be reproduced by those methods.


\section*{Acknowledgments}

The authors thank Marc Henneaux for useful discussions. This work is partially
funded by FONDECYT grants 1020629, 1040921, 3040026, 1051056 and 1061291. O.M.
was supported by PUCV through the program Investigador Joven 2006. The generous
support to CECS by Empresas CMPC is also acknowledged. CECS is a Millenium
Science Institute and is funded in part by grants from Fundaci\'{o}n Andes and
the Tinker Foundation.

\appendix

\section{Supersymmetric extension of AdS$_{5}$, $SU(2,2|N)$}

The supersymmetric extension of the AdS group in five dimensions, $SO(2,4)$, is
the super unitary group $SU(2,2\left|N\right.)$ \cite {Nahm,Strathdee},
containing supermatrices of unit superdeterminant which leave invariant the
(real) quadratic form
\begin{equation}
q=\theta ^{*\alpha }G_{\alpha \beta }\theta ^{\beta }+z^{*r}g_{rs}z^{s}\qquad
(\alpha =1,\ldots ,4;\;\;r=1,\ldots ,N)\,. \label{q form}
\end{equation}
Here $\theta ^{\alpha }$ are complex Grassman numbers (with complex conjugation
defined as $\left( \theta ^{\alpha }\theta ^{\beta }\right) ^{*}=\theta
^{*\beta }\;\theta ^{*\alpha }$), and $G_{\alpha \beta }$ and $ g_{rs}$ are
Hermitean matrices, antisymmetric and symmetric respectively, which can be
chosen as
\begin{equation}
G_{\alpha \beta }=i\left( \Gamma _{0}\right) _{\alpha \beta }\,,\qquad \qquad
g_{rs}=\delta _{rs}\,.
\end{equation}
The bosonic sector of this supergroup is
\begin{equation}
SU(2,2)\otimes SU(N)\otimes U(1)\subset SU(2,2|N)\,,
\end{equation}
where the AdS group is present on the basis of the isomorphism $ SU(2,2)\simeq
SO(2,4).$ Therefore, the generators of $su(2,2\left| N\right. ) $ algebra are
\begin{equation}
\begin{array}{lll}
so(2,4):\quad & \mathbf{J}_{AB}=\left( \mathbf{J}_{ab,}\mathbf{J}_{a}\right)
\,,\qquad \medskip & \left( A,B=0,\ldots ,5\right) \,, \\
su(N): & \mathbf{T}_{\Lambda }\,,\medskip & \left( \Lambda =1,\ldots
,N^{2}-1\right) \,, \\
\mbox{SUSY}: & \mathbf{Q}_{s}^{\alpha },\;\mathbf{\bar{Q}}_{\alpha
}^{s}\,,\medskip & \left( \alpha =1,\ldots ,4;\;s=1,\ldots ,N\right) \,, \\
u(1): & \mathbf{Z}\,, &
\end{array}
\end{equation}
where $\eta _{AB}=$ diag $\left( -,+,+,+,+,-\right) ,$ and AdS rotations and
translations are $\mathbf{J}_{ab}$ and$\;\mathbf{J}_{a}\equiv \mathbf{J}%
_{a5}\;(a,b=0,\ldots ,4)$. The dimension of this superalgebra is $\Delta
=N^{2}+8N+15$. For $N=1$, the generators $\mathbf{T}_{\Lambda }$ are absent,
and the bosonic sector is given by AdS$_5\otimes u(1)$ algebras.

A representation of the superalgebra acting in $\left( 4+N\right) $%
-dimensional superspace $\left( \theta ^{\alpha },y^s\right) $ is given by the
$\left( 4+N\right) \times \left( 4+N\right) $ supermatrices
\begin{equation}
\begin{tabular}{llllll}
$\mathbf{J}_{AB}=$ & $\left(
\begin{array}{cc}
\frac{1}{2}\,\left( \Gamma _{AB}\right) _{\alpha }^{\beta } & 0 \\
0 & 0
\end{array}
\right) \,,\medskip \quad $ & $\mathbf{Q}_{s}^{\alpha }=$ & $\left(
\begin{array}{cc}
0 & 0 \\
-\delta _{s}^{r}\delta _{\beta }^{\alpha } & 0
\end{array}
\right) \,,\quad $ & $\mathbf{Z}=$ & $\left(
\begin{array}{cc}
\frac{i}{4}\,\delta _{\alpha }^{\beta } & 0 \\
0 & \frac{i}{N}\,\delta _{r}^{s}
\end{array}
\right) \,,$ \\
$\mathbf{T}_{\Lambda }=$ & $\left(
\begin{array}{cc}
0 & 0 \\
0 & \left( \tau _{\Lambda }\right) _{r}^{s}
\end{array}
\right) \,,$ & $\mathbf{\bar{Q}}_{\alpha }^s=$ & $\left(
\begin{array}{cc}
0 & \delta _{r}^{s}\delta _{\alpha }^{\beta } \\
0 & 0
\end{array}
\right) \,,$ &  &
\end{tabular}
\end{equation}
where the $4\times 4$ matrices $\Gamma _{AB}$ are defined as
\begin{equation}
\Gamma _{ab}=\frac{1}{2}\,\left[ \Gamma _{a},\Gamma _{b}\right] \,,\qquad
\qquad \Gamma _{a5}=\Gamma _{a}\,,
\end{equation}
$\Gamma _{a}$ are the Dirac matrices in five dimensions with the signature $%
\left( -,+,+,+,+\right) $, and $\tau _{\Lambda }$ are anti-Hermitean generators
of $su(N)$ acting in $N$-dimensional space $y^s$.

>From the given representation of supermatrices it is straightforward to find
the explicit form of the corresponding Lie algebra. The commutators of the
bosonic generators $\mathbf{J}_{AB}$, $\mathbf{T}_{\Lambda }$ and $\mathbf{Z}$
closes the algebra $su(2,2)\otimes su(N)\otimes u(1)$, while the supersymmetry
generators transforms as spinors under AdS and as vectors under $su(N)$,
\begin{equation}
\begin{array}{ll}
\left[ \mathbf{J}_{AB},\mathbf{Q}_s^{\alpha }\right] =-\frac{1}{2}\,\left(
\Gamma _{AB}\right) _{\beta }^{\alpha }\,\mathbf{Q}_s^{\beta }\,,\qquad \qquad
\medskip & \left[ \mathbf{T}_{\Lambda },\mathbf{Q}_s^{\alpha }\right] =\left(
\tau _{\Lambda }\right) _s^r\,\mathbf{Q}_r^{\alpha
}\,, \\
\left[ \mathbf{J}_{AB},\mathbf{\bar{Q}}_{\alpha }^s\right]
=\frac{1}{2}\,\mathbf{\bar{Q}}_{\beta }^s\,\left( \Gamma _{AB}\right) _{\alpha
}^{\beta }\,, & \left[ \mathbf{T}_{\Lambda }, \mathbf{\bar{Q}}_{\alpha
}^s\right] =- \mathbf{\bar{Q}}_{\alpha }^r\,\left( \tau _{\Lambda }\right)
_r^s\,,
\end{array}
\end{equation}
and they carry $u(1)$ charges,
\begin{equation}
\left[ \mathbf{Z},\mathbf{Q}_s^{\alpha }\right] =-i\,\left( \frac{1}{4}
-\frac{1}{N}\right) \,\mathbf{Q}_s^{\alpha }\,,\qquad \qquad \left[
\mathbf{Z},\mathbf{\bar{Q}}_{\alpha }^s\right] =i\,\left( \frac{1}{4}-
\frac{1}{N}\right) \,\mathbf{\bar{Q}}_{\alpha }^s\,.  \label{U(1) Charges}
\end{equation}
The anticommutator of the supersymmetry generators has the form
\begin{equation}
\left\{ \mathbf{Q}_s^{\alpha }\mathbf{,\bar{Q}}_{\beta }^r\right\} =
\frac{1}{4}\,\delta _s^r\,\left( \Gamma ^{AB}\right) _{\beta }^{\alpha
}\,\mathbf{J}_{AB}-\delta _{\beta }^{\alpha }\,\left( \tau ^{\Lambda }\right)
_s^r\,\mathbf{T}_{\Lambda }+i\,\delta _{\beta }^{\alpha }\,\delta
_s^r\,\mathbf{Z}\,.
\end{equation}
>From these expressions it is clear that for $N=4$ the $U(1)$ generator
$\mathbf{Z}$ becomes a central charge and the algebra becomes a central
extension of $PSU(2,2\left\vert 4\right. )$.

An invariant third rank tensor, completely symmetric in bosonic and
antisymmetric in fermionic indices, can be constructed as
\begin{equation}
ig_{KLM}\equiv \left\langle \mathbf{G}_{K}\mathbf{G}_{L}\mathbf{G}
_M\right\rangle =\frac{1}{2}\,\mbox{Str\thinspace }\left[ \left(
\mathbf{G}_K\mathbf{G}_L+\left( -\right) ^{\varepsilon _K\varepsilon _L}
\mathbf{G}_L\mathbf{G}_K\right) \mathbf{G}_M\right] \, , \label{3}
\end{equation}
with the following non-vanishing components:
\begin{equation}
\begin{array}{lll}
g_{\left[ AB\right] \left[ CD\right] \left[ EF\right] } & = & -\frac{1}{2}%
\,\varepsilon _{ABCDEF}\,,\medskip \\
g_{\Lambda _{1}\Lambda _{2}\Lambda _{3}} & = & -\gamma _{\Lambda _{1}\Lambda
_{2}\Lambda _{3}}\,,\medskip \\
g_{\left[ AB\right] \left( _{r}^{\alpha }\right) \left( _{\beta }^{s}\right) }
& = & -\frac{i}{4}\,\left( \Gamma _{AB}\right) _{\beta }^{\alpha }\delta
_{r}^{s}\,,\medskip \\
g_{\Lambda \left( _{r}^{\alpha }\right) \left( _{\beta }^{s}\right) } & = & -
\frac{i}{2}\,\delta _{\beta }^{\alpha }\left( \tau _{\Lambda }\right)
_{r}^{s}\,,
\end{array}
\qquad \qquad
\begin{array}{lll}
g_{z\left[ AB\right] \left[ CD\right] } & = & -\frac{1}{4}\,\eta _{\left[
AB\right] \left[ CD\right] }\,,\medskip \\
g_{z\Lambda _{1}\Lambda _{2}} & = & -\frac{1}{N}\,\gamma _{\Lambda
_{1}\Lambda _{2}}\,,\medskip \\
g_{z\left( _{r}^{\alpha }\right) \left( _{\beta }^{s}\right) } & = &
\frac{1}{2}\,\left( \frac{1}{4}+\frac{1}{N}\right) \delta _{\beta }^{\alpha
}\delta
_{r}^{s}\,,\medskip \\
g_{zzz} & = & \,\frac{1}{N^{2}}-\frac{1}{4^{2}}\,,
\end{array}
\end{equation}
Here $\eta _{\left[ AB\right] \left[ CD\right] }\equiv \eta _{AC}\,\eta
_{BD}-\eta _{AD}\,\eta _{BC}\,$ and $\gamma _{\bar{K}\bar{L}}$ are the Killing
metrics of $SO(2,4)$ and $SU(N)$, respectively. The symmetric third rank
invariant tensor for $su(N)$ is  $\gamma _{\Lambda _{1}\Lambda _{2}\Lambda
_{3}}\equiv \frac{1}{2i}\,$Tr%
$_{N}\left( \left\{ \tau _{\Lambda _{1}},\tau _{\Lambda _{2}}\right\} \tau
_{\Lambda _{3}}\right) $, and the $\Gamma $-matrices are normalized so that
\begin{equation}
\mbox{Tr}_{4}\left( \Gamma _{a}\,\Gamma _{b}\,\Gamma _{c}\,\Gamma _{d}\,\Gamma
_{e}\right) =-4i\,\varepsilon _{abcde}\,,\qquad \left( \varepsilon
^{abcde5}\equiv \varepsilon ^{abcde},\quad \varepsilon ^{012345}=1\right) \,.
\end{equation}

Splitting the generators as $\mathbf{G}_{K}=\left( \mathbf{G}_{\bar{K}},
\mathbf{Z}\right) $, it can be seen that the invariant tensor for $SU(2,2|N)$
fulfills the conditions: (\emph{i}) $g_{\bar{K}\bar{L}z}$ is invertible, and
(\emph{ii}) $g_{\bar{K}zz}$ vanishes. Then, as shown in
\cite{Miskovic-Troncoso-Zanelli}, it is easy to identify generic configuration
and those satisfying eq. (\ref{canonical1}, \ref{canonical2}) are canonical.

In the special case $N=4$, the invariant tensor $g_{KLM}$ of $SU(2,2\left|
4\right.)$ simplifies to:
\begin{equation}
\begin{tabular}{llllll}
$g_{\left[ AB\right] \left[ CD\right] \left[ EF\right] }$ & $=$ & $
-\frac{1}{2}\,\varepsilon _{ABCDEF}\,,$ & $g_{\Lambda \left( _{r}^{\alpha
}\right) \left( _{\beta }^{s}\right) }$ & $=$ & $-\frac{i}{2}\,\delta _{\beta
}^{\alpha }\left( \tau _{\Lambda }\right) _{r}^{s}\,,\medskip $ \\
$g_{\Lambda _{1}\Lambda _{2}\Lambda _{3}}$ & $=$ & $-\gamma _{\Lambda
_{1}\Lambda _{2}\Lambda _{3}}\,,\medskip $ & $g_{z\bar{K}\bar{L}}$ &
$=$ & $-\frac{1}{4}\,\gamma _{\bar{K}\bar{L}}\,,\medskip $ \\
$g_{\left[ AB\right] \left( _{r}^{\alpha }\right) \left( _{\beta }^{s}\right)
}$ & $=$ & $-\frac{i}{4}\,\left( \Gamma _{AB}\right) _{\beta }^{\alpha }\delta
_{r}^{s}\,,\qquad \qquad \qquad $ &  &  &
\end{tabular}
\end{equation}
with $g_{zzz}=0$ and  $\gamma _{\bar{K}\bar{L}}$ is the Killing metric for
$PSU(2,2\left| 4\right.)$ . In this case, the dimension of the group is
$\Delta=63$.



\end{document}